\documentclass[reprint,onecolumn,showkeys,showpacs,preprintnumbers,nofootinbib,amsmath,amssymb,aps,prd,longbibliography,10pt]{revtex4}
\usepackage{array,multirow,graphicx}
\usepackage{amsmath,amssymb,dsfont}
\usepackage{dcolumn}
\usepackage{bm}
\usepackage[colorlinks,citecolor=blue,urlcolor=blue,linkcolor=blue]{hyperref}

\begin{document}
\title{Lorenz Gauge Fixing of $f(T)$ Teleparallel Cosmology}
\author{W. El Hanafy$^{1,3}$}%
\email{waleed.elhanafy@bue.edu.eg}
\author{G.G.L. Nashed$^{1,2,3}$}%
\email{nashed@bue.edu.eg}
\affiliation{$^{1}$Centre for Theoretical Physics, The British University in Egypt, P.O. Box 43, El Sherouk City, Cairo 11837, Egypt}
\affiliation{$^{2}$Mathematics Department, Faculty of Science, Ain Shams University, Cairo 11566, Egypt}
\affiliation{$^{3}$Egyptian Relativity Group (ERG), Cairo University, Giza 12613, Egypt}
\begin{abstract}
In teleparallel gravity, we apply Lorenz type gauge fixing to cope with redundant degrees of freedom in the vierbein field. This condition is mainly to restore the Lorentz symmetry of the teleparallel torsion scalar. In cosmological application, this technique provides standard cosmology, turnaround, bounce or $\Lambda$CDM as separate scenarios. We reconstruct the $f(T)$ gravity which generates these models. We study the stability of the solutions by analyzing the corresponding phase portraits. Also, we investigate Lorenz gauge in the unimodular coordinates, it leads to unify a nonsingular bounce and standard model cosmology in a single model, where crossing the phantom divide line is achievable through a finite-time singularity of Type IV associated with a de Sitter fixed point. We reconstruct the unimodular $f(T)$ gravity which generates the unified cosmic evolution showing the role of the torsion gravity to establish a healthy bounce scenario.
\end{abstract}

\maketitle
\section{Introduction}\label{Sec:1}
The recent detection of the gravitational waves \cite{PhysRevLett.115.181101} by the laser interferometer gravitational-wave observatory (LIGO) ensures that gravitation is a geometrical phenomena as formulated in the general relativity (GR) theory, which attributes gravitation to the curvature of the spacetime itself. Universality of the free falling makes the geometrical description of the GR possible. In this sense, geometry replaces force and geodesic replaces equation of motion \cite{2013tegr.book.....A}. An alternative theory, but equivalent, that allows gravitation to be a gauge theory is the teleparallel equivalent of general relativity (TEGR). However, it attributes gravitation to torsion, which acts as a lorentz force in the equations of motion.

Although, both theories predict the same set of the field equations, they realize gravitation in two different perceptions. In $4$-dimensional GR, all geometrical quantities are built from ten components of a symmetric metric $g_{\mu\nu}$. In $4$-dimentional TEGR, all geometrical quantities are built from sixteen components of a tetrad vector fields $e_{a}^{\mu}$. In the former, there is a unique torsionless linear connection (Levi-Civita) represents both gravitational and inertial effects, so that for a particular local frame gravitation and inertial effects could exactly compensate each other. On the contrary, in TEGR, there is a unique curvatureless linear connection (Weitzenb\"{o}ck), so that the covariant derivative of the tetrad with respect to this connection identically vanishes, which defines the teleparallelism nature. The same derivative is not covariant under local Lorentz group, which makes all the local lorentz invariants rotate freely at every point of the spacetime \cite{Maluf:2012yn}. Since, both theories give the same set of the field equations for the same degrees of freedom, the extra six components in the tetrad vector field are related to the local Lorentz transformation and do not represent a physical field.

Although, both connections see the spacetime with a different resolution, there are some bridges to link these two extreme geometries. One of these is that the Levi-Civita curvature (Ricci) scalar $R$ and the Weitzenb\"{o}ck (teleparallel) torsion scalar $T$ are geometrically differ by a total derivative term. Therefore, when the lagrangian formulation are carried out for both, we get the same field equations. In spite of the scalar R is invariant under both diffeomorphism and local Lorentz groups, the total derivative is not invariant under the latter, which subsequently makes the teleparallel torsion scalar not a Lorentz invariant \cite{Maluf:2012yn}. This essential difference alters the equivalence when both theories have been extended by replacing $R$ and $T$ by arbitrary functions $f(R)$ and $f(T)$. The field equations of the later lack invariance under local Lorentz transformation \cite{Li:2010cg,Sotiriou:2010mv}.

It has been shown that the total derivative can be made to vanish, in order to eliminate the local Lorentz dependance in the $f(T)$ solution in the case of the spherical symmetry \cite{Nashed:2015AHEP}. In this paper, we extend the procedure to the cosmological applications by considering the homogeneity and isotropy symmetries. We examine two cases: when the total derivative term has been made to vanish or to be a constant. Surprisingly, we get the big bang or $\Lambda$CDM models, where the constant total derivative term acts just as the cosmological constant. This tells that restoring the lorentz symmetry enforces the two extremists $R$ and $T$ to nearly compensate each other.

We organize the work as follows: In Sec. \ref{Sec:2}, we review the concept of the linear connection and its role to realize different geometrical relations. The Ricci scalar of Riemman geometry and the teleparallel torsion scalar of the Weitzenb\"{o}ck geometry differ by a total derivative term. We require this term to be a constant to restore the Lorentz symmetry of the teleparallel torsion scalar at this particular case. We show that this condition is an analog to the Lorenz gauge fixing in electromagnetism. In Sec. \ref{Sec:3}, we reconstruct the $f(T)$ gravity which generates these scenarios. In Sec. \ref{Sec:4}, we show that the $f(T)$ cosmology provides a one dimensional autonomous system. Therefore, the stability of the dynamical behavior can be determined completely by the stability of the fixed points and the asymptotic behavior of the phase portraits. Their analysis clarifies the nature of the constant, which acts exactly as the cosmological constant. Also, it shows that the solution can give rise to standard, bounce, turnaround or $\Lambda$CDM as separate cosmological models. In Sec. \ref{Sec:5}, we apply Lorenz gauge in the unimodular coordinates, whereas the cosmological constant appears as a constant of integration and does not gravitate. However, we find the Lorenz gauge gives rise to a cosmological model can unify the bounce and the standard cosmologies. The phase portrait analysis shows that the universe is nonsingular origin at Minkowski space, perform a bouncing behavior preceding the standard cosmology phase. We show that the crossing of phantom divide line, as required for bouncing models, is completely secured through traversable Type IV singularities at the de Sitter fixed point during crossing between phantom and nonphantom regimes. Also, we evaluate the $f(T)$ which generates this dynamical behavior, showing the role of the torsion gravity to avoid the conventional problems of the bouncing cosmology. In Sec. \ref{Sec:6}, we conclude the work.
\section{Vielbein space and teleparallelism}\label{Sec:2}
In an $n$-space equipped with $n$-independent vielbein vector fields $e_{a}$ ($a=1,\cdots, n$) defined globally on a smooth manifold $\mathcal{M}$ we call this a vielbein space. Let $e_{a}{^{\mu}}$ $(\mu = 1, ..., n)$ be the coordinate components of the $a$-th vector field $e_{a}$, where Greek and Latin indices are constrained by the Einstein summation convention\footnote{The Latin indices are $SO(3,1)$ or Lorentz indices, and the Greek are space-time indices.}. In this natural space one can construct many linear affine connections \cite{Wanas:1986,Wanas:1999wc,Wanas:2016,Wanas:2017105}, for each there is a different associated geometry of the same space. However, one can go to two extreme cases:
\subsection{Weitzenb\"{o}ck connection}\label{Sec:2.1}
The covariant components $e_{a \mu}$ of $e_{a}$ are given via the relations $e_{a}{^{\mu}}e^{a}{_{\nu}}=\delta^{\mu}_{\nu}$ and $e_{a}{^{\mu}}e^{b}{_{\mu}}=\delta^{b}_{a}$, where $\delta$ is the Kronecker tensor. It can be shown that the determinant $e\equiv \det (e_{a}{^{\mu}})\neq 0$ as a consequence of the independence of $e_{a}$. On the vielbein space $(M,\,e_{a})$, there exists a unique nonsymmetric linear connection, namely Weitzenb\"{o}ck connection, with respect to which the vielbein vector fields $e_{a}$ are parallel. This connection is given by $\Gamma^{\lambda}{_{\mu\nu}}\equiv e_{a}{^\lambda}\partial_{\mu}e^{a}{_\nu}$, and is characterized by the teleparallelism condition, that is $\nabla^{(\Gamma)}_{\nu}e_{a}{^{\mu}}\equiv 0$, where the operator $\nabla^{(\Gamma)}_{\nu}$ is the covariant derivative with respect to the Weitzenb\"{o}ck connection.

The non-commutation of an arbitrary vector fields $V_{a}$ is given by
\begin{equation}
\nabla^{(\Gamma)}_{\nu}\nabla^{(\Gamma)}_{\mu}V_{a}{^{\alpha}} - \nabla^{(\Gamma)}_{\mu}\nabla^{(\Gamma)}_{\nu}V_{a}{^{\alpha}}= R(\Gamma){^{\alpha}}{_{\epsilon\mu\nu}} V_{a}{^{\epsilon}} + T(\Gamma)^{\epsilon}{_{\nu\mu}} \nabla^{(\Gamma)}_{\epsilon} V_{a}{^{\alpha}},
\end{equation}
where $R(\Gamma)^{\alpha}{_{\epsilon\mu\nu}}$ and $T(\Gamma)^{\epsilon}{_{\nu\mu}}$ are, respectively, the curvature and the torsion tensors of Weitzenb\"{o}ck connection. The above non-commutation formula along with the teleparallelism condition force the curvature tensor $R(\Gamma)^{\alpha}_{~~\mu\nu\sigma}$ to vanish identically, while the torsion tensor $T(\Gamma)_{a\mu\nu}=\partial_{\mu}e_{a\nu}-\partial_{\nu}e_{a\mu}$ does not vanish. In this geometry one can define the contortion tensor via $K_{\alpha \mu \nu}=\frac{1}{2}\left(T_{\nu\alpha\mu}+T_{\alpha\mu\nu}-T_{\mu\alpha\nu}\right)$, where $T_{\lambda\mu\nu}=e^{a}{_{\lambda}}T_{a\mu\nu}$.
\subsection{Levi-Civita connection}\label{Sec:2.2}
The vielbein fields can define a metric tensor on $M$ by the simple relation $g_{\mu \nu} \equiv \eta_{ab}e^{a}{_{\mu}}e^{b}{_{\nu}}$, with inverse metric $g^{\mu \nu} = \eta^{ab}e_{a}{^{\mu}}e_{b}{^{\nu}}$. Thus one can go straightforward to construct the full picture of Riemannian geometry, where the symmetric Levi-Civita linear connection associated with $g_{\mu\nu}$ is given by $\overcirc{\Gamma}{^{\alpha}}{_{\mu\nu}}= \frac{1}{2} g^{\alpha \sigma}\left(\partial_{\nu}g_{\mu \sigma}+\partial_{\mu}g_{\nu \sigma}-\partial_{\sigma}g_{\mu \nu}\right)$. As is well known the non-commutation relation in Riemannan geometry context implies a vanishing torsion tensor $T(\overcirc{\Gamma})^{\epsilon}{_{\nu\mu}}=0$, but a non-vanishing curvature tensor $R(\overcirc{\Gamma})^{\alpha}{_{\epsilon\mu\nu}}\neq 0$.
\subsection{Geometrical identities bridges}\label{Sec:2.3}
Indeed we deal with exactly one space. However, Levi-Civita and Weitzenb\"{o}ck connection can see this space with different resolution. The former represents an extreme picture with a vanishing torsion tensor, while the later represent another extreme with a vanishing curvature tensor. Interestingly, one can find possible links between these two extremes. For example, in the view of the teleparallelism condition, the Weitzenb\"{o}ck connection is metric, i.e. $\nabla^{(\Gamma)}_{\sigma}g_{\mu\nu}\equiv 0$. Also, the contortion can be seen as the difference between Weitzenb\"{o}ck and Levi-Civita connections $K^{\alpha}{_{\mu\nu}} \equiv \Gamma^{\alpha}_{~\mu\nu} - \overcirc{\Gamma}{^{\alpha}}_{\mu\nu}=e_{a}{^{\alpha}}~ \nabla^{(\overcirc{\Gamma})}_{\nu}e^{a}{_{\mu}}$, where the covariant derivative $\nabla^{(\overcirc{\Gamma})}_{\sigma}$ is with respect to the Levi-Civita connection. One more useful links in this context is the following identity
\begin{equation}\label{identity}
    eR(\overcirc{\Gamma})\equiv -eT(\Gamma)+2\partial_{\mu}(eT^{\mu}),
\end{equation}
where $e=\sqrt{-g}=\det\left({e}_\mu{^a}\right)$, also the teleparallel scalar torsion $T$ is given in the literature in many forms
\begin{eqnarray}\label{Tsc1}
 T \equiv \Sigma^{abc}T_{abc}= \frac{1}{4}T^{abc}T_{abc}+\frac{1}{2}T^{abc}T_{acb}-T^{a}T_{a},
\end{eqnarray}
where $T_{a}=T^{b}{_{ba}}$ and the superpotential $\Sigma^{abc}$ is given by
\begin{equation}\label{superpotential}
    \Sigma^{abc}=\frac{1}{4}\left(T^{abc}+T^{bac}-T^{cab}\right)+\frac{1}{2}\left(\eta^{ac}T^{b}-\eta^{ab}T^{c}\right)
\end{equation}
is skew symmetric in the last pair of indices. As seen from identity (\ref{identity}), the teleparallel torsion scalar density is equivalent to the curvature scalar density up to a total derivative term. This helped to develop a gravitational theory equivalent to the general theory of relativity (TEGR). This has been achieved by using the right hand side of (\ref{identity}) instead of the curvature scalar density in Einstein-Hilbert action where the total derivative term does not contribute in the variation. This can be shown by considering the action of TEGR
\begin{eqnarray}
\nonumber {\cal S}_{TEGR}&=&\frac{-1}{2\kappa^2}\int e T d^4x \equiv \frac{1}{2\kappa^2} \int [e R+2\partial_{\mu}(eT^{\mu})]d^4x \equiv \frac{1}{2\kappa^2}\int eRd^4x+2\int_{\partial M} T^{\mu} n_\mu d^3x\\
&\equiv& \frac{1}{2\kappa^2}\int \sqrt{-g} R d^4x = {\cal S}_{GR},
\end{eqnarray}
where $\kappa^2=8\pi G/c^4$, $c$ is the speed of light in vacuum and $G$ is the Newtonian's gravitational constant. Although the apparent equivalence between the GR and the TEGR on the level of the field equations, the difference runs deep on the lagrangian level. Therefore, their extensions
\begin{equation}\label{eq}
 \int \sqrt{-g} f(R)d^4x\neq \int e f(T)d^4x.
 \end{equation}
The reason for the inequality in Eq. (\ref{eq}) is that the boundary term $T^{\mu} n_\mu$ in the nonlinear case does not behave as a surface term.
\subsection{Lorenz gauge fixing}\label{Sec:2.4}
For a spatial homogeneous and isotropic universe, the metric is in Friedmann-Robertson-Walker (FRW) form
\begin{equation}\label{FRW-metric}
ds^2=c^2 dt^{2}-a(t)^{2}\delta_{ij} dx^{i} dx^{j},
\end{equation}
where $a(t)$ is the scale factor of the universe. We assume the natural units $k_{B}=c=\hbar=1$, the Ricci scalar is given as
\begin{equation}\label{Ricciscalar}
    R(t)=-6\left[\frac{\ddot{a}}{a}+\left(\frac{\dot{a}}{a}\right)^{2}\right]=-6\left(\dot{H}+2 H^{2}\right),
\end{equation}
where $H\equiv \dot{a}^2/a^2$ is the Hubble parameter and the dot denotes the differentiation with respect to the cosmic time. On the contrary, the vierbein corresponding to the FRW metric (\ref{FRW-metric}) may take the diagonal form
\begin{equation}\label{FRW-tetrad}
{e_{\mu}}^{a}=\textmd{diag}(1,a(t),a(t),a(t)).
\end{equation}
Then, the teleparallel torsion scalar is given by
\begin{equation}\label{Torsionscalar}
    T(t)=-6\left(\frac{\dot{a}}{a}\right)^{2}=-6 H^{2}.
\end{equation}
In the geometric identity (\ref{identity}), the Ricci scalar in the left hand side is a Lorentz invariant, so the right-hand side is invariant too. Since, the total derivative term is not a Lorentz invariant, the teleparallel torsion scalar is not a Lorentz invariant \cite{Li:2010cg,Sotiriou:2010mv}. Although this serious difference does not affect the TEGR formulation to be in conflict with the Lorentz symmetry, it has a crucial role in the $f(T)$ formulation. Here, we impose a gauge fixing condition to restore the Lorentz symmetry of the teleparallel torsion scalar by taking
\begin{equation}\label{Lorenz-gauge}
    \partial_{\mu}(eT^{\mu})=-2\Lambda,
\end{equation}
where $\Lambda$ is some constant with units [length]$^{-2}$. As is well known that the teleparallel gravity can be seen as a gauge theory under translation group, where the torsion tensor $T^{a}{_{\mu\nu}}$ represents the field strength. In this sense, we find the condition (\ref{Lorenz-gauge}) is similar to the Lorenz gauge of the gauge field, i.e. $\partial_{\mu} A^{\mu}=0$, in electromagnetism. Substituting (\ref{Ricciscalar}), (\ref{Torsionscalar}) and (\ref{Lorenz-gauge}) into (\ref{identity}), we write
\begin{equation}\label{divergence}
    \partial_{\mu}(eT^{\mu})\equiv\frac{1}{2}e\left[R+T\right]
    =-6\left[\frac{\ddot{a}}{a}+2\left(\frac{\dot{a}}{a}\right)^{2}\right]=-2\Lambda.
\end{equation}
In the following, we investigate the two cases, $\Lambda=0$ as a special choice, and the more general case $\Lambda\neq 0$, which may provide a physical interpretation of this constant.
\subsubsection{The case: $\Lambda=0$}
Solving (\ref{divergence}) for the scale factor where $\Lambda= 0$, we get
\begin{equation}\label{scalefactor0}
    a(t)=\left[3H_{0}(t-t_{i})\right]^{1/3},
\end{equation}
where $H_{0}\equiv H(t_{0})$ and $t_{i}$ are constants of integration. It is clear that the obtained model belongs to the standard cosmology and the model has an initial singularity at $t=t_{i}$, where $t_{i}$ denotes the time of the initial singularity. By taking the initial conditions $a(0)=0$ and $a_{0}\equiv a(t_{0})=1$, where $t_{0}$ denotes the present time, the constants can be determined to be $t_{i}=0$ and $H_{0}=\frac{1}{3t_{0}}$. Using (\ref{scalefactor0}), we evaluate Hubble function and the teleparallel torsion scalar as
\begin{equation}\label{Hubble0}
    H(t)=\frac{1}{3(t-t_{i})}, \quad T(t)=-\frac{2}{3(t-t_{i})^{2}}.
\end{equation}
Notably, the solution belongs to the standard big bang model when the matter has a linear equation of state $p_m=(\gamma-1)\rho_m$, i.e. $a\propto t^{\frac{2}{3\gamma}}$, when the matter field is stiff matter ($\gamma=2$).
\subsubsection{The case: $\Lambda\neq 0$}
Solving (\ref{divergence}) for the scale factor where $\Lambda\neq 0$, we get
\begin{equation}\label{scalefactor1}
    a(t)=\frac{3}{2\sqrt{\Lambda}}\left(c_{2} e^{-\sqrt{\Lambda}t}-c_{1}e^{\sqrt{\Lambda}t}\right)^{1/3}.
\end{equation}
Thus, the Hubble rate and the teleparallel torsion scalar, respectively, are
\begin{eqnarray}
    H(t)&=&\frac{\sqrt{\Lambda}}{3}\left[\frac{c_{1}e^{\sqrt{\Lambda}t}+c_{2}e^{-\sqrt{\Lambda}t}}
    {c_{1}e^{\sqrt{\Lambda}t}-c_{2}e^{-\sqrt{\Lambda}t}}\right],\label{Hubble1} \\ T(t)&=&-\frac{2\Lambda}{3}\left[\frac{c_{1}e^{\sqrt{\Lambda}t}+c_{2}e^{-\sqrt{\Lambda}t}}
    {c_{1}e^{\sqrt{\Lambda}t}-c_{2}e^{-\sqrt{\Lambda}t}}\right]^{2}.\label{Tsc2}
\end{eqnarray}
\section{Reconstruction of $f(T)$ gravity}\label{Sec:3}
We consider the simple generalization of the TEGR action by introducing an arbitrary function of the teleparallel torsion scalar $f(T)$,
\begin{equation}\label{action}
{\cal S}=\int d^{4}x~ e\left[\frac{1}{2 \kappa^2}f(T)+L_{m}\right],
\end{equation}
where $L_{m}$ is the lagrangian of the matter. The variation of the action (\ref{action}) with respect to the vierbein gives \cite{BF09,L10}
\begin{equation}\label{field_eqns}
\frac{1}{e} \partial_\mu \left( e S_a^{\verb| |\mu\nu} \right) f^{\prime}-e_a^\lambda  T^\rho_{\verb| |\mu \lambda} S_\rho^{\verb| |\nu\mu}f^{\prime}+S_a^{\verb| |\mu\nu} \partial_\mu T f^{\prime\prime}
+\frac{1}{4} e_a^\nu f=\frac{{\kappa}^2}{2} e_a^\mu \mathfrak{T}_\mu^{\verb| |\nu},
\end{equation}
where $f':=\frac{df}{dT}$ and $f'':=\frac{d^{2}f}{dT^2}$, and ${\mathfrak{T}_{\mu}}^{\nu}$ is the usual energy-momentum tensor of matter fields. Assume that the material-energy tensor is taken for a perfect fluid
\begin{equation}\label{matter}
\mathfrak{T}_{\mu\nu}=\rho u_{\mu}u_{\nu}+p(u_{\mu}u_{\nu}-g_{\mu\nu}),
\end{equation}
where $u^{\mu}$ is the fluid 4-velocity, $\rho_{m}$ and $p_{m}$ are the energy density and pressure of the fluid in its rest frame. Using the field equations (\ref{field_eqns}) for the matter (\ref{matter}), the modified Friedmann equations for the $f(T)$-gravity read \cite{BF09,L10}
\begin{eqnarray}
  \rho_m &=& \frac{1}{2\kappa^2}\left[f(T)+12 H^2 f'\right], \label{FR1T}\\
  p_m &=& \frac{-1}{2\kappa^2}\left[f(T)+4(3H^2+\dot{H})f'-48\dot{H}H^2 f''\right].\label{FR2T}
\end{eqnarray}
The $f(T)$ teleparallel generalization has received attentions in the last decade either in astrophysical or cosmological applications \cite{Shirafuji:1996,Nashed:2002NCimB,Nashed2002,Nashed:2010ApSS,Wei:2011mq,Sadjadi:2012xa,Bamba:2012vg,Nashed:2012ChPhL,Cardone:2012xq,
Nashed:2013PhRvD,Nashed:2014lva,Camera:2013bwa,Odintsov:2015AnPhy,Myrzakulov:2015,Sharif:2015,Rodrigues:2015JCAP,
Capozziello:2015PRD,Karami,Chen1,Oikonomou:2016PhRvD,Nunes:2016JCAP,Krssak:2016CQGra,Bamba:2016,ElHanafy:2016a,ElHanafy:2015,Ruggiero,ElHanafy:2016b,
ElHanafy:2016c,Nashed:2017,Awad:2017JCAP}. A review on $f(T)$ teleparallel gravity can be found in Ref.~\cite{Cai:2016RPPh}. We constrain the matter fluid to be perfect with a linear equation-of-state, where $\gamma=1$ for dust, $\gamma=4/3$ for radiation and $\gamma=2$ for stiff matter.
\subsection{The case: $\Lambda=0$}
Using the chain rule $f^{\prime} = \dot{f}/\dot{T},~ f^{\prime\prime} = \left(\dot{T} \ddot{f}-\ddot{T} \dot{f}\right)/\dot{T}^{3} $, and substituting from (\ref{Hubble0}) into the field equations (\ref{FR1T}) and (\ref{FR2T}), we write the density and pressure as follows
\begin{eqnarray}
  \rho_m &=& \frac{1}{2\kappa^2}\left[f(t)-\frac{H}{\dot{H}}\dot{f}(t)\right]=\frac{1}{2\kappa^2}\left[f(t)+(t-t_{i})\dot{f}(t)\right], \label{FR1}\\[2pt]
\nonumber p_m&=&-\frac{1}{2\kappa^2}\left[f(t)-\left(\frac{H}{\dot{H}}-\frac{\ddot{H}}{3\dot{H}^{2}}\right)\dot{f}(t)
-\frac{1}{3\dot{H}}\ddot{f}(t)\right]\\
&=&-\frac{1}{2\kappa^2}\left[f(t)+3(t-t_{i})\dot{f}(t)+(t-t_{i})^{2}\ddot{f}(t)\right].\label{FR2}
\end{eqnarray}
For a perfect fluid with a linear equation-of-state the continuity equation of the matter reads
\begin{equation}\label{continuity}
    \dot{\rho}_m+3\gamma H \rho_m=0.
\end{equation}
The integration of (\ref{continuity}) gives
\begin{equation}\label{density00}
\rho_m=\rho_{m,0}e^{-3\gamma \int H dt}=\rho_{m,0}\left[3H_{0}(t-t_{i})\right]^{-\gamma},
\end{equation}
where $\rho_{m,0}\equiv \rho(t_{0})$ is a constant. Combining (\ref{FR1}) and (\ref{density00}) and by solving for $f(t)$ we get
\begin{equation}\label{ft0}
    f(t)=\frac{c_{1}}{t-t_{i}}-\frac{2\rho_{0}\kappa^{2}/(\gamma-1)}{\left[3H_{0}(t-t_{i})\right]^{\gamma}},
\end{equation}
where $c_{1}$ is constant of integration. Using the inverse relation of (\ref{Hubble0}) we can express the time in terms of the torsion scalar, the usual form of $f(T)$ can be obtained as
\begin{equation}\label{fT0}
    f(T)=c'_{1}\sqrt{T}-\frac{2\rho_{0}\kappa^{2}}{\gamma-1}\sqrt{T^{\gamma}}.
\end{equation}
To be consistent with the solution, the fluid has to be stiff matter $\gamma=2$, which produces a linear $f(T)$ theory or GR. This is verified in the perturbation level of the $f(T)$ theory, when the equations of motion of the fluctuations produces the GR profile when the speed of sound $c_{s}^{2}=\frac{\delta p_{m}}{\delta \rho_{m}}=1$, i.e. when the matter is stiff \cite{Cai:2011tc}.
\subsection{The case: $\Lambda\neq 0$}
Substituting from (\ref{Hubble1}) into the field equations (\ref{FR1}) and (\ref{FR2}), we write the density and pressure as follows
\begin{eqnarray}
    \rho_{m}&=&\frac{1}{2\kappa^2}\left[f(t)+\frac{1}{4}\left(\lambda^{-1} e^{2\sqrt{\Lambda}t}-\lambda e^{-2\sqrt{\Lambda}t}\right)\dot{f}(t)\right]\label{density1}\\
\nonumber p_{m}&=&-\frac{1}{2\kappa^2}\left[f(t)+\frac{3}{4}\left(\lambda^{-1} e^{2\sqrt{\Lambda}t}-\lambda e^{-2\sqrt{\Lambda}t}\right)\dot{f}(t)\right.\\
&&\left. +\frac{1}{4\Lambda}\left(\lambda^{-1/2}e^{\sqrt{\Lambda}t}-\lambda^{1/2}e^{-\sqrt{\Lambda}t}\right)^{2}
    \ddot{f}(t)\right],\label{pressure1}
\end{eqnarray}
where $\lambda=c_{2}/c_{1}$. Following the same procedure of Eqs. (\ref{FR1}--\ref{ft0}), we evaluate the $f(t)$ for the case of $\Lambda\neq 0$ as
\begin{equation}\label{f1(t)}
    f(t)=(c_{1}e^{2\sqrt{\Lambda}t}-c_{2})^{-\gamma}\left[c_{3}\mathcal{H}_{1}+c_{4}e^{(\gamma+2)\sqrt{\Lambda}t}
    \mathcal{H}_{2}\right],
\end{equation}
where $\mathcal{H}$ are hypergeometric functions, that are defined as follows
\begin{eqnarray}
\nonumber  \mathcal{H}_{1} &=& \textmd{hypergeom}\left(\left[-\gamma,1-\frac{\gamma}{2}\right],\left[-\frac{\gamma}{2}\right],
\lambda^{-1}e^{2\sqrt{\Lambda}t}\right),\\
\nonumber \mathcal{H}_{2} &=& \textmd{hypergeom}\left(\left[2,1-\frac{\gamma}{2}\right],\left[2+\frac{\gamma}{2}\right],\lambda^{-1}e^{2\sqrt{\Lambda}t}\right).
\end{eqnarray}
Thus the corresponding $f(T)$-theory which generates a constant boundary term is given by (\ref{f1(t)}). Using the inverse relation of (\ref{Hubble1}), one can reexpress (\ref{f1(t)}) in its regular $f(T)$ form as below
\begin{equation}\label{f1(T)}
    f(T)=(3T+2\Lambda)^{\frac{1}{2}+\frac{\gamma}{4}}\left[c'_{3}\mathcal{LP}_{1}+c'_{4}\mathcal{LQ}_{1}\right]
\end{equation}
where
\begin{eqnarray}
\nonumber  \mathcal{LP}_{1} &=& \textmd{LegenderP}\left(\frac{-\gamma}{2},1+\frac{\gamma}{2},\sqrt{\frac{-T}{2\Lambda/3}}\right), \\
\nonumber \mathcal{LQ}_{1} &=& \textmd{LegenderQ}\left(\frac{-\gamma}{2},1+\frac{\gamma}{2},\sqrt{\frac{-T}{2\Lambda/3}}\right).
\end{eqnarray}
At $T=-\frac{2\Lambda}{3}$, the $f(T)$ drops to zero. In fact, this case is unphysical, at $T=-\frac{2\Lambda}{3}$ the universe is exactly de Sitter fixed point, which cannot be reached in a finite time. This point will be clarified in the following when we discuss the stability of the solution.
\section{Stability of the solutions}\label{Sec:4}
In the curvature based modified gravity, e.g. $f(R)$, one can see that the dependence on the higher derivatives of $H$ is unavoidable. On the contrary, the torsion based modified gravity, e.g. $f(T)$, shows dependence on the Hubble parameter and its first derivative only, which enables to write the modified Friedmann equations as a one dimensional autonomous system, i.e. $\dot{H}=\mathcal{F}(H)$ \cite{Awad:2013PRD,Bamba:2016,Awad:2017JCAP,ElHanafy:2017xsm,AHNS:2016}. The plot of this relation in ($H$, $\dot{H}$) phase space is called the phase portrait, while the solution $H(t)$ is the phase trajectory. This nice feature allows one of the basic techniques to study the dynamical evolution of the Friedmann system, that is the stability of the phase portrait fixed points.
\subsection{One dimensional autonomous system}\label{Sec:4.1}
We apply the above argument in the case of the $f(T)$ cosmology. So it is convenient to write Eqs. (\ref{FR1T}) and (\ref{FR2T}) in terms of $H$, using (\ref{Torsionscalar}), we write \cite{AHNS:2016}
\begin{eqnarray}
  \rho_m &=& \frac{1}{2\kappa^2}\left[f(H)-H f_{H}\right], \label{FR1H}\\
  p_m &=& \frac{-1}{2\kappa^2}\left[f(H)-H f_{H}-\frac{1}{3}\dot{H} f_{HH}\right],\label{FR2H}
\end{eqnarray}
where $f_{H}:=\frac{d~f}{dH}$ and $f_{HH}:=\frac{d^{2}f}{dH^2}$. Holding the linear equation of state $p=(\gamma-1)\rho$, we write \cite{AHNS:2016}
\begin{equation}\label{phasetrajectory}
    \dot{H}=3\gamma \left[\frac{f(H)-H f_{H}}{f_{HH}}\right]=\mathcal{F}(H),
\end{equation}
which provides the phase portrait for any $f(T)$ theory. It is easy to show that (\ref{phasetrajectory}) reproduces the GR limit by setting $f(H)=-6H^2$, then the phase portrait reduces to
\begin{equation}\label{GR-phport}
    \dot{H}=-\frac{3}{2}\gamma H^2.
\end{equation}
It worths to mention that the middle plot in Fig. \ref{Fig: phasespace} can be produced by the above phase portrait for the stiff matter case $\gamma=2$.
\subsection{Fixed points of phase portraits}\label{Sec:4.2}
\subsubsection{The case: $\Lambda=0$}
\begin{figure}
\centering
\includegraphics[scale=.4]{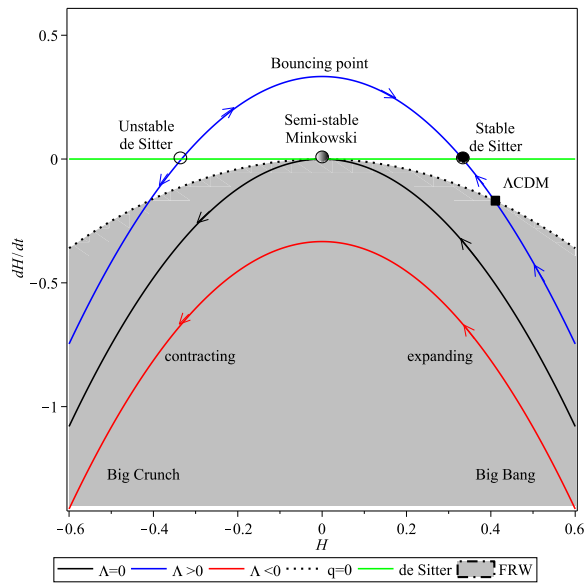}
\caption{phase-space: (a) $\Lambda>0$, (b) $\Lambda=0$, (c) $\Lambda<0$.}
\label{Fig: phasespace}
\end{figure}

For the solution (\ref{scalefactor0}), we have the phase portrait
\begin{equation}\label{phaseportrait0}
    \dot{H}=-3H^{2}.
\end{equation}
Confronting the above equation with the GR phase portrait (\ref{GR-phport}) implies that $\gamma=2$. This is the case of the standard model of cosmology, when the universe is dominated by a stiff matter. The phase portrait plot is given in Fig. \ref{Fig: phasespace}, where the curve asymptotically $H\to \pm \infty$ evolves towards a finite-time singularity of Type I, where the time of the singularity \cite{Awad:2013PRD}
\begin{equation}
\nonumber t_s=-\int_{H_0}^{\infty}\frac{dH}{\dot{H}}=\frac{1}{3H_{0}}.
\end{equation}

On the other hand, the curve passes through the Minkowskian origin ($H=0$, $\dot{H}=0$). In fact, this is a semi-stable fixed point and the universe takes an infinite time to reach that point, i.e.
\begin{equation}
\nonumber t=\int_{H_0}^{0}\frac{dH}{\dot{H}}\to \infty.
\end{equation}
Therefore, the fixed point split the portrait into two separate patches: The $H>0$ patch, at which the universe begins with a Type I singularity (big bang) followed by a decelerated expansion phase evolving towards Minkowski space at $H=0$. The $H>0$ patch, at which the universe has no initial singularity, since it begins with a fixed point, then it is followed by a decelerated contraction phase evolving asymptotically as $H\to -\infty$ towards a future finite-time singularity of Type I (big crunch).
\subsubsection{The case: $\Lambda\neq 0$}
For the solution (\ref{scalefactor1}), we have the phase portrait
\begin{equation}\label{phaseportrait1}
    \dot{H}=-3H^{2}+\frac{\Lambda}{3}.
\end{equation}
The corresponding phase portrait is given in Fig. \ref{Fig: phasespace}. As clear from the plot, the constant $\Lambda$ is in fact the cosmological constant. Recalling the geometric identity (\ref{identity}), the following question raises:\textit{ Why do the Ricci and the teleparallel torsion scalars just about to oppose each other?}

We discuss two possible subclasses, when the cosmological constant is negative or positive.

\paragraph{For $\Lambda<0$.} The phase portrait is shifted vertically downwards, and the universe is one patch, where the transition from expanding ($H>0$) to contracting ($H<0$) deceleration phase is valid. Therefore, the universe interpolates between two finite-time singularities of Type I: From big bang at $H\to \infty$ to big crunch at $H\to -\infty$.\\

\paragraph{For $\Lambda>0$.} The phase portrait is shifted vertically upwards, the portrait has two fixed points at $H=\pm \frac{\sqrt{\Lambda}}{3}$. Since the time required to reach any of these fixed points is infinite, they split the phase portrait into three separate patches: (\textbf{i}) The first is when $H>\frac{\sqrt{\Lambda}}{3}$, in this patch the universe begins with an initial singularity of Type I (big bang) at time
$$t_s=-\int_{H_0}^{\infty}\frac{dH}{\dot{H}}=\frac{\mathfrak{R}\left(\textmd{arctanh} \left(\frac{3 H_0}{\sqrt{\Lambda}}\right)\right)}{\sqrt{\Lambda}}.$$
However, the plot cuts the zero acceleration curve at $\dot{H}=-H^2$, which identifies the transition from deceleration to acceleration at $H_{de}=\sqrt{\frac{\Lambda}{6}}$. Notably, the value of the constant $\Lambda$ can be determined by knowing the precise value of the Hubble parameter $H_{de}$ at the transition. After, the universe evolves to a fixed point, which represents a stable de Sitter space. It can be shown that the universe has no future singularity, since the time required to reach that point is infinite. In this patch the solution gives a $\Lambda$CDM model. (\textbf{ii}) The second is when $-\frac{\sqrt{\Lambda}}{3}<H<\frac{\sqrt{\Lambda}}{3}$, in this patch the universe interpolates between two fixed points, so its age $-\infty<t<\infty$. As shown in the plot the universe evolves from contraction $H<0$ to expansion $H>0$, where $\dot{H}>0$ at $H=0$. This gives a nonsingular bounce universe where the system begins at an unstable de Sitter universe at $H=-\frac{\sqrt{\Lambda}}{3}$ in a phantom regime and evolves towards a stable de Sitter phase at $H=\frac{\sqrt{\Lambda}}{3}$. (\textbf{iii}) The third patch is when $H<-\frac{\sqrt{\Lambda}}{3}$, in this patch the universe has no initial singularity, since it begins with a fixed point. As clear From Fig. \ref{Fig: phasespace}, the universe has an early accelerated contraction, then it enters a decelerated phase of contraction. Finally, it evolves towards a future finite time singularity of Type I (big crunch), where the big crunch time can be calculated as shown before.

In conclusion, for a positive cosmological constant, the model predicts a nonsingular bouncing model when $-\frac{\sqrt{\Lambda}}{3}<H<\frac{\sqrt{\Lambda}}{3}$, while it performs the standard model of cosmology when $H>\frac{\sqrt{\Lambda}}{3}$, in this patch the universe begins with an initial singularity of Type I (big bang), then it evolves as a dark fluid that interpolates between dust and de Sitter at late time, just as in $\Lambda$CDM cosmology. The bouncing patch solves the initial singularity problem and even the trans-Planckian problem of inflationary scenarios. on the other hand, the dark fluid patch explains the late transition from deceleration to acceleration. However, these two patches cannot be exist in a unified model, since they are separated by a fixed point.
\section{Unifying bounce and standard cosmology}\label{Sec:5}
In this section, we turn our discussion into the unimodular gravity. We show that the bounce solution can be unified with the standard cosmology in a single framework by applying Lorenz gauge fixing of the unimodular teleparallel gravity.
\subsection{Unimodular conditions}\label{Sec:5.1}
For the FRW metric (\ref{FRW-metric}), we turn to the unimodular coordinates by introducing a new time variable $\tau$ \cite{Gomez:2016PRD}
\begin{equation}\label{unimodular-condition}
    d\tau = a(t)^{3}dt.
\end{equation}
Thus, the metric reads
\begin{equation}\label{Unimodular-metric}
ds^2=a(\tau)^{-6} d\tau^{2}-a(\tau)^{2}\delta_{ij} dx^{i} dx^{j},
\end{equation}
and the corresponding tetrad becomes
\begin{equation}\label{unimodular-tetrad}
{e_{\mu}}^{a}=\textmd{diag}\left(a(\tau)^{-3},a(\tau),a(\tau),a(\tau)\right),
\end{equation}
where $|e|=\sqrt{-g}=1$, which satisfies the unimodular condition. It has been shown that the unimodular condition reduce the symmetry of the spacetime by one degree of freedom, since the field equations become invariant under the subgroup of the diffeomorphism, that is the transverse diffeomorphism. Therefore, the unimodular condition does not add a further constraint to reduce the degrees of freedom \cite{daRochaNeto:2011ir}.
\subsection{Lorenz-like gauge fixing}\label{Sec:5.2}
Now we reconstruct the unimodular Ricci scalar
\begin{equation}\label{uni.Ricciscalar}
    \mathcal{R}(\tau)=-6a(\tau)^{4}\left[a\left(\frac{d^{2}a}{d\tau^{2}}\right)+4\left(\frac{da}{d\tau}\right)^{2}\right]
                 =-6a(\tau)^{6}\left(\dot{\mathcal{H}}+5 \mathcal{H}^{2}\right).
\end{equation}
where the dot denotes the derivative with respect to $\tau$, and $\mathcal{H}\equiv \frac{1}{a(\tau)}\left(\tfrac{da}{d\tau}\right)$ plays the role of the Hubble parameter. Also, the reconstructed unimodular teleparallel torsion scalar reads
\begin{equation}\label{uni.Torsionscalar}
    \mathcal{T}(\tau)=-6 a(\tau)^{4}\left(\frac{da}{d\tau}\right)^{2}=-6 a(\tau)^{6} \mathcal{H}^{2}.
\end{equation}
Similar to Sec. \ref{Sec:2.4}, we consider the identity (\ref{identity}), therefore
\begin{equation}\label{uni.divergence}
    \mathcal{R}+\mathcal{T} =-2\Lambda,
\end{equation}
where $\Lambda$ is some constant with units [length]$^{-2}$. Using the identity (\ref{identity}), the above equation gives rise to a Lorenz-like gauge condition
\begin{equation}\label{uni.Lorenz-condition}
    \nabla^{(\overcirc{\Gamma})}_{\mu}\mathcal{T}^{\mu}=-\Lambda.
\end{equation}
For the special case $\Lambda=0$ of the unimodular universe model (i.e. $|e|=1$), Eq. (\ref{uni.Lorenz-condition}) reduces to Lorenz condition
$$\partial_{\mu}\mathcal{T}^{\mu}=0.$$
Substituting (\ref{uni.Ricciscalar}) and (\ref{uni.Torsionscalar}) into (\ref{uni.divergence}), we obtain
\begin{equation}\label{divergence1}
    a(\tau)^{4}\left[5\left(\frac{da}{d\tau}\right)^{2}+a(\tau) \frac{d^{2}a}{d\tau^{2}}\right]
    =a(\tau)^{6}\left[\dot{\mathcal{H}}+6\mathcal{H}\right]
    =\frac{\Lambda}{3}.
\end{equation}
Solving the above differential equation for the scale factor $a(\tau)$, we get
\begin{equation}\label{uni.scale-factor}
    a(\tau) = \left(\Lambda \tau^{2}-6c_{1} \tau + 6c_{2}\right)^{1/6},
\end{equation}
where $c_{1}$ and $c_{2}$ are constants of integration. In the case of $\Lambda=0$, the solution is a big bang model. For simplicity, we introduce new parameters $a_{B}=\left(6c_{2}-9\frac{c_{1}^{2}}{\Lambda}\right)^{1/6}$, $\beta=\frac{\Lambda}{a_{B}^{6}}$ and $\tau_{B}=3\frac{c_{1}}{\Lambda}$; then the scale factor (\ref{uni.scale-factor}) reads
\begin{equation}\label{uni.scale-factor1}
    a(\tau)=a_{B}\left[\beta~ \left(\tau-\tau_{B}\right)^{2}+1\right]^{1/6}.
\end{equation}
Since $-\infty < \tau < \infty$, the above form gives a non-singular bouncing scale factor. We take $a_{B}$ is its value at the bouncing time $\tau_{B}$, and $\beta$ describes how fast the bounce occurs. Using (\ref{unimodular-condition}) and (\ref{uni.scale-factor1}), the relation between the unimodular time and the cosmic time can be expressed by
\begin{equation}\label{cosmic-time}
    t(\tau) = t_0+\frac{\ln \left[\sqrt{\beta}~(\tau-\tau_{B})+\sqrt{\beta(\tau-\tau_B)^2+1}\right]}{a_{B}^{3}\sqrt{\beta}},
\end{equation}
where $t_{0}$ is a constant of integration.
\subsection{The Cosmological Phase Portrait}\label{Sec:3.3}
\begin{figure}
\centering
\includegraphics[scale=.45]{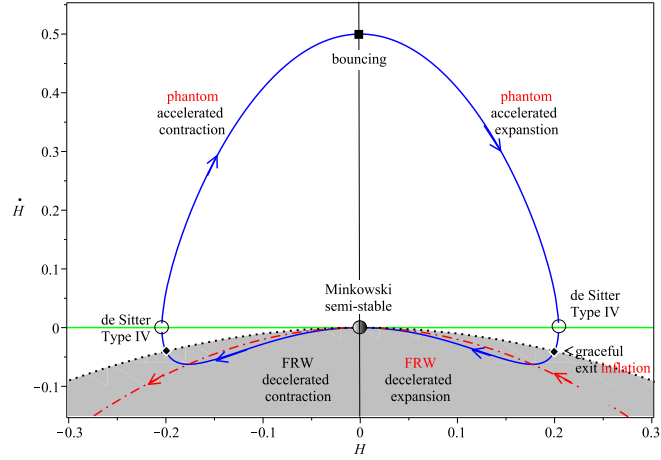}
\caption{The phase portrait of Eq. (\ref{Uni-phase-portrait}). The plot shows the non-singular behavior Minkowskian origin ($\mathcal{H},\mathcal{\dot{H}}$) $\equiv$ ($0,0$) is at $\tau\rightarrow \pm\infty$}
\label{Fig:phase-portrait}
\end{figure}

Using Eq. (\ref{uni.scale-factor1}), we evaluate the Hubble parameter
\begin{equation}\label{Uni-Hubble}
    \mathcal{H}=\frac{\beta (\tau-\tau_{B})}{3(\beta (\tau-\tau_{B})^{2}+1)}.
\end{equation}
Consequently, we express the time $\tau$ in terms of $\mathcal{H}$ as
\begin{equation}\label{uni-time}
    \tau(\mathcal{H})_{\pm}=\frac{\beta(1+6\beta \mathcal{H}\tau_{B}) \pm \sqrt{\beta(\beta-36 \mathcal{H}^{2})}}{6\beta \mathcal{H}}.
\end{equation}
This enabled us to construct the model phase portrait by expressing $\dot{\mathcal{H}}$ as a function of $\mathcal{H}$
\begin{equation}\label{Uni-phase-portrait}
    \dot{\mathcal{H}}_{\pm}=-\frac{6\mathcal{H}^{2}\sqrt{\beta(\beta-36 \mathcal{H}^{2})}}{\mp \beta +\sqrt{\beta(\beta-36 \mathcal{H}^{2})}},
\end{equation}
where the plus (minus) subscript denotes the branch $\dot{\mathcal{H}}>0$ ($\dot{\mathcal{H}}<0$), respectively. The above equation represents a one-dimensional autonomous system, which can be represented geometrically as shown in Fig. (\ref{Fig:phase-portrait}). The ($\mathcal{H}-\dot{\mathcal{H}}$) phase space corresponding to Eq. (\ref{Uni-phase-portrait}) can help to reveal the dynamical evolution of Universe in a clear and a transparent analysis with no need to solve the system \cite{Awad:2013PRD}.\\
\subsubsection{The Minkowskian origin}
Since fixed points dominate the dynamics of these type of dynamical systems, we identify the fixed points by finding the roots of Eq. (\ref{Uni-phase-portrait}), i.e. we solve $\dot{\mathcal{H}}=0$. This determines the fixed points at
$$\mathcal{H}=0,~\pm \frac{\sqrt{\beta}}{6}.$$
The graphical representation of the phase portrait, Fig. \ref{Fig:phase-portrait}, shows that the universe is trapped within the Hubble interval $|\mathcal{H}|<\frac{\sqrt{\beta}}{6}$. At the origin, the fixed point $\mathcal{H}=0$ represents Minkowskian space. The universe evolves away from the origin, if an initial Hubble $\mathcal{H}_{1}<0$. But it evolves toward the Minkowskian fixed point, if the initial Hubble $\mathcal{H}_{1}>0$. Therefore, we classify the Minkowskian origin as a semi-stable fixed point. Since at a fixed point, i.e. $\dot{\mathcal{H}}=0$, the universe needs an infinite time to approach it. Therefore, the time required to reach the Minkowskian fixed point $\mathcal{H}_{0}$ on the $\dot{\mathcal{H}}_{-}$ branch
$$\tau=\int_{\mathcal{H}_{1}\neq 0}^{0}\frac{d\mathcal{H}}{\dot{\mathcal{H}}_{-}}=\infty.$$
It is natural, now, to choose an initial Hubble $\mathcal{H}_{1}<0$, just about the Minkowskian origin, where there is no past initial singularity.
\subsubsection{Decelerated contraction}
Following the phase space flow, the universe begins within region II. Since $\mathcal{H}<0$ and $q>0$ at that region, the universe experiences a decelerated contraction phase. This phase ends when the phase portrait (\ref{Uni-phase-portrait}), of the $\dot{\mathcal{H}_{-}}$ branch, cuts the zero acceleration curve at which $q=0$ or equivalently $\dot{\mathcal{H}_{-}}=-\mathcal{H}^{2}$. So, at $\mathcal{H}<0$, the intersection is at $\mathcal{H}_{2}=- \frac{\sqrt{6~\beta}}{15}.$
This also ensures that the parameter $\beta$ is positive. The time interval from the initial Hubble $\mathcal{H}_{1}$ to the transition point $\mathcal{H}_{2}$ can be given by
$$\tau=\int_{\mathcal{H}_{1}}^{\mathcal{H}_{2}}\frac{d\mathcal{H}}{\dot{\mathcal{H}}_{-}}=-\frac{3\sqrt{6}\mathcal{H}_{1}
+\sqrt{\beta}+\sqrt{\beta-36\mathcal{H}_{1}^{2}}}{6\sqrt{\beta}\mathcal{H}_{1}}.$$
Thus, the total time during the decelerated contraction phase is as given above.\\

\subsubsection{Accelerated contraction}
By crossing the transition point, the universe enters region I, and a new phase of an accelerating contraction has begun. However, the universe is still in the $\dot{\mathcal{H}}_{-}$ branch, we take the interval to reach the next fixed point $\mathcal{H}_{3}=-\frac{\sqrt{\beta}}{6}$, which represents a de Sitter space. Usually, the time required to reach a fixed point is an infinite. However, this is true for an equation of state $p_{\textmd{eff}}(\mathcal{H})$, and where the pressure is continuous and differentiable. On the contrary, when the pressure is not differentiable, i.e. $dp_{\textmd{eff}}/d\mathcal{H}$ is not continuous, we have two possible cases: The first is when the discontinuity of $dp_{\textmd{eff}}/d\mathcal{H}$ is finite, but the time to reach a fixed point is also infinite. The second case when $dp_{\textmd{eff}}/d\mathcal{H}$ is infinite discontinuous, it is the only possible option to reach a fixed point in a finite time. From Eq. (\ref{FR2H}), the last case implies the divergence of quantity $d\dot{\mathcal{H}}/d \mathcal{H}$. Therefore, the only possible case that the universe can approach the fixed point $\mathcal{H}_{3}$ in a finite time, when the following conditions are fulfilled \cite{Awad:2013PRD}:
\begin{itemize}
  \item [(i)] $\lim_{\mathcal{H}\rightarrow \mathcal{H}_{3}} \dot{\mathcal{H}}_{\pm}=0$,
  \item [(ii)] $\lim_{\mathcal{H}\rightarrow \mathcal{H}_{3}} d\dot{\mathcal{H}}_{\pm}/d\mathcal{H}=\pm \infty$,
  \item [(iii)] $\tau=\int_{\mathcal{H}}^{\mathcal{H}_{3}}d\mathcal{H}/\dot{\mathcal{H}}_{\pm}<\infty$.
\end{itemize}
In conclusion, the only possible case to reach a fixed point in a finite time is to have an infinite slope of the phase portrait at the fixed point. In the general relativistic picture, the divergence of $d\dot{\mathcal{H}}/d \mathcal{H}$ leads to a divergence of the speed of sound $dp_{m}/d\rho_{m}=c_{s}^2$, so the solution will not be causal.

It is clear that the above three conditions are fulfilled at the fixed point $\mathcal{H}_{3}$. Moreover, the time interval between the transition $\mathcal{H}_{2}$ and the de Sitter fixed point $\mathcal{H}_{3}$ is given by
$$\tau=\int_{\mathcal{H}_{2}}^{\mathcal{H}_{3}}\frac{d\mathcal{H}}{\dot{\mathcal{H}}_{-}}=\frac{-2+\sqrt{6}}{2\sqrt{\beta}}.$$
This makes the universe to stay at in the contracting $\dot{\mathcal{H}}_{-}$ branch an interval of time
$$\tau=\int_{\mathcal{H}_{1}}^{\mathcal{H}_{3}}\frac{d\mathcal{H}}{\dot{\mathcal{H}}_{-}}=-\frac{6\mathcal{H}_{1}
+\sqrt{\beta}+\sqrt{\beta-36\mathcal{H}_{1}^{2}}}{6\sqrt{\beta}\mathcal{H}_{1}}.$$
\subsubsection{Phantom crossing I (Type IV singularity)}
It has been shown that the above mentioned conditions are not sufficient to let the universe to cross the fixed point $\mathcal{H}_{3}$ at a finite time. But we need the phase portrait to be a double valued function to cross the phantom divide line . As clear from Fig. (\ref{Fig:phase-portrait}), the phase portrait has not only an infinite slope at $\mathcal{H}_{3}$, but also it is a double valued function. This shows that the de Sitter space is accessible, and the crossing from $w_{\textmd{eff}}>-1$ to $w_{\textmd{eff}}<-1$ is possible. Such a double valued function often appears when there is a first-order phase transition.

At the de Sitter fixed point, $\mathcal{H}_{3}=-\frac{\sqrt{\beta}}{6}$, it can be shown that the scale factor, the Hubble parameter and its first derivative have finite values. Consequently, $\rho_{\textmd{eff}}$ and $p_{\textmd{eff}}$ are finite as well. However, from Eq. (\ref{Uni-phase-portrait}), we find the second derivative of the Hubble parameter
\begin{equation}\label{TypeIV}
\ddot{\mathcal{H}}=\dot{\mathcal{H}}\left(\frac{d\dot{\mathcal{H}}}{d\mathcal{H}}\right)=\frac{72 \mathcal{H}^{2}\left[\beta^{3}-54\mathcal{H}^{2}\beta^{2}+\left(\beta^2-36\mathcal{H}^{2}\beta\right)^{3/2}\right]}
{\left(\beta+\sqrt{\beta^{2}-36\mathcal{H}^{2}\beta}\right)^{3}}.
\end{equation}
In general, for a fixed point $\mathcal{H}_{f}$ to be reached in a finite time, we have $d\dot{\mathcal{H}}_{\pm}/d\mathcal{H}\rightarrow \pm \infty$ as $\mathcal{H}\rightarrow \mathcal{H}_{f}$. Consequently, we expect the second derivative of $\ddot{\mathcal{H}}$ to diverge, and the de Sitter space in this case represents a finite-time singularity of type IV \cite{AHNS:2016,ElHanafy:2017xsm}. In particular, in this model, Eq. (\ref{TypeIV}) shows that the second derivative at the de Sitter fixed point $\mathcal{H}_{3}=-\frac{\sqrt{\beta}}{6}$ is diverging. This type finite-time singularity shows interesting features, that is the most mild, phenomenologically, at which the geodesic are complete. In addition to the allowance of type IV finite time singularity at de Sitter phase, also it has no effect on the observed amplitude of the gravitational wave. However, it affects only the dynamics of inflation, which allows the universe to pass smoothly through this type of singularities \cite{Kleidis:2016vmd}.
\subsubsection{Bouncing point and trans-Plankian problem}
In Fig. (\ref{Fig:phase-portrait}), the phase portrait cuts the $\dot{\mathcal{H}}$ axis. At the intersection point ($\mathcal{H}_{B}=0$, $\dot{\mathcal{H}}>0$), the universe bounces from an accelerated contraction ($\mathcal{H}<0$) to an accelerated expansion phase ($\mathcal{H}>0$), while $\dot{\mathcal{H}}>0$ before and after the bouncing point. On the $\dot{\mathcal{H}}_{+}$ branch, the time required after crossing the de Sitter point $\mathcal{H}_{3}$ to reach the bouncing point $\mathcal{H}_{B}=0$ is given by
$$\tau=\int_{\mathcal{H}_{3}}^{\mathcal{H}_{B}}\frac{d\mathcal{H}}{\dot{\mathcal{H}}_{+}}=\frac{1}{\sqrt{\beta}}.$$
\begin{figure}
\centering
\includegraphics[scale=.38]{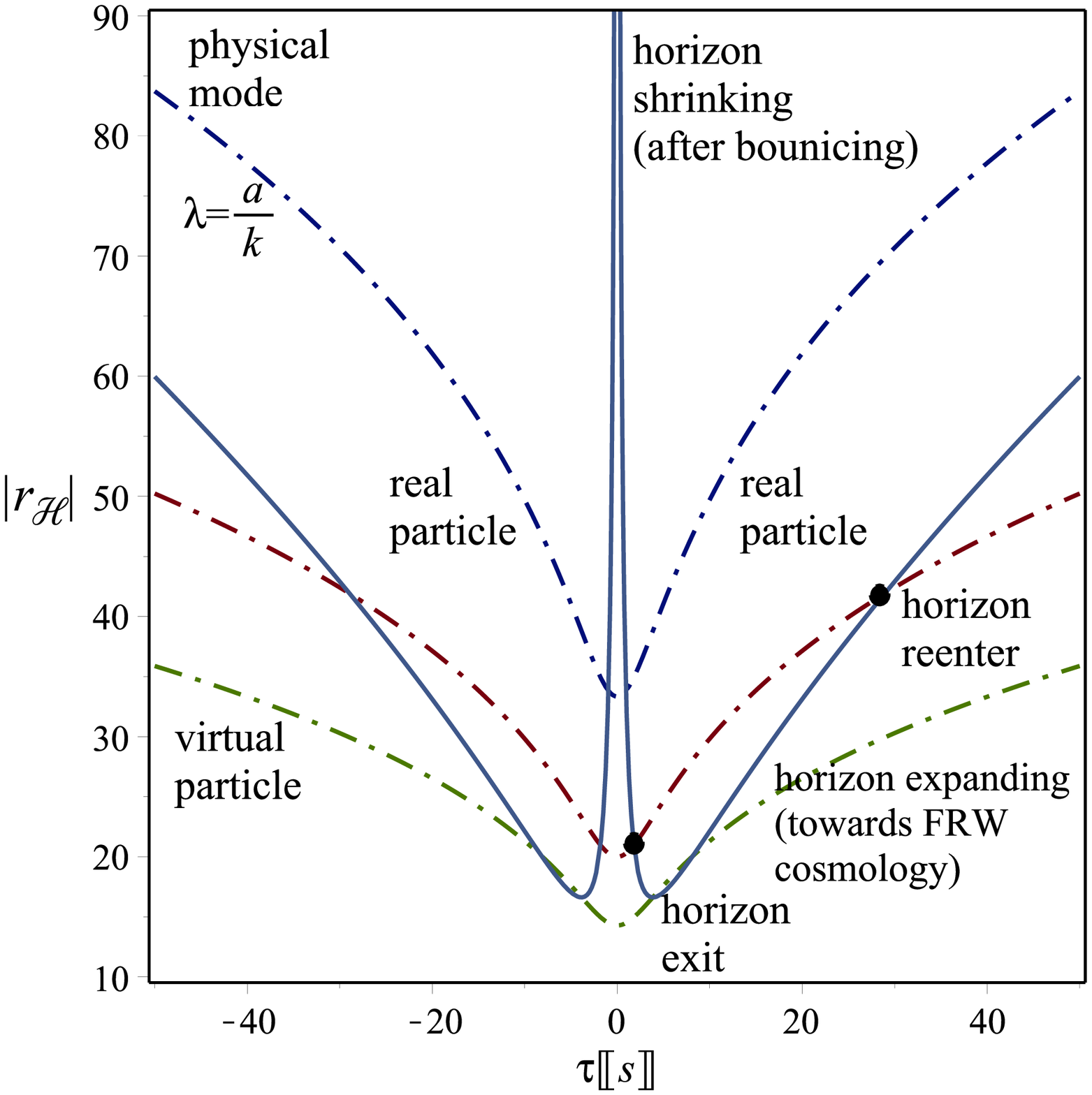}
\caption{The evolution of the comoving hubble radius (horizon): Eq. (\ref{hubble-radius}).}
\label{Fig:hubble-radius}
\end{figure}
In inflationary models, the comoving hubble radius (horizon), $r_{\mathcal{H}}=\frac{1}{a(\tau)\mathcal{H}}$, is infinite at the initial singularity as the initial scale factor $a(\tau=0)=0$. Since all wavelengthes $\lambda \ll r_{H}$, all the primordial quantum fluctuations are subhorizon\footnote{Equivalently, the comoving wave number $k$ is at subhorizon scales as $k \gg a\mathcal{H}=\frac{1}{r_{\mathcal{H}}}$.}. On the contrary, the Hubble horizon is infinite at the bouncing point as $\mathcal{H}_{B}(\tau=0)=0$. Consequently, we expect all wavelengthes to be subhorizon at the bouncing time. In inflationary models, the scale factor is growing exponentially and the Hubble is constant. Therefore, the horizon is exponentially shrinking during inflation, allowing some modes to exit the horizon and freeze out. By the end of inflation, the horizon begins to expand allowing the freezed modes to reenter the horizon at superhorizon scales and propagate as particles. In this bouncing model, the comoving hubble radius
\begin{equation}\label{hubble-radius}
    r_{\mathcal{H}}=\frac{3\left(\beta\tau^{2}+1\right)^{5/6}}{\beta\tau}.
\end{equation}
The evolution of the Hubble horizon after bounce is represented by Fig. (\ref{Fig:hubble-radius}). The plot shows that the how fast Hubble horizon shrinks after bounce to a minimal value allowing some modes to exit the horizon, then it expands allowing the freezed modes to reenter the a horizon. So the bouncing mechanism gains the benefits of inflationary models, it remains to examine the conservation of the primordial fluctuation between horizon exit and reenter. This should be treated through a field theory at the perturbation level.
\subsubsection{Accelerated expansion}
After bouncing, the universe enters a new phase of an accelerating expansion as $\mathcal{H}>0$ and $\dot{\mathcal{H}}>0$. We take the final point of this phase at the next fixed point (de Sitter space) $\mathcal{H}_{4}=\frac{\sqrt{\beta}}{6}$. At this fixed point, we find that the three mentioned conditions, at the fixed point $\mathcal{H}_{3}$, are also fulfilled. Therefore, the time interval to complete this phase is finite, and can be given by
$$\tau=\int_{\mathcal{H}_{B}}^{\mathcal{H}_{4}}\frac{d\mathcal{H}}{\dot{\mathcal{H}}_{+}}=\frac{1}{\sqrt{\beta}}.$$
This shows that the transition from contraction to expansion about the bouncing point is symmetric.
\subsubsection{Phantom crossing II (Type IV singularity)}
In addition to the three conditions, one can find that the phase portrait (\ref{Uni-phase-portrait}) is a double valued function. This qualifies the universe for a second finite time transition of the phantom divide line from $w_{\textmd{eff}}<-1$ to $w_{\textmd{eff}}>-1$ at the fixed point $\mathcal{H}_{4}$. Just as in the phantom crossing I at the fixed point $\mathcal{H}_{3}$, one can classify the fixed point $\mathcal{H}_{4}$ as a finite time singularity of type IV. Consequently, we expect the geodesics to be complete, and the amplitudes of the gravitational waves are not affected by crossing this fixed point.
\subsubsection{Graceful exit inflation}
After crossing the phantom divide line, the universe, again, evolves on the $\dot{\mathcal{H}}_{-}$ branch, see Fig. \ref{Fig:phase-portrait}. However, it remains in an accelerated expansion phase until cutting the zero acceleration curve $\dot{\mathcal{H}}_{-}=-\mathcal{H}^{2}$. We identify the intersection point $\mathcal{H}_{5}=\frac{\sqrt{6~\beta}}{15}$. This allows a time interval
$$\tau=\int_{\mathcal{H}_{4}}^{\mathcal{H}_{5}}\frac{d\mathcal{H}}{\dot{\mathcal{H}}_{-}}=\frac{-2+\sqrt{6}}{2\sqrt{\beta}}.$$
During this time interval, the universe evolves effectively in a nonphantom regime, $-1<w_{\textmd{eff}}<-1/3$, just as in the canonical inflation. Combining the time intervals from bouncing point to de Sitter, then to transition point $\mathcal{H}_{5}$, the total time that universe stays in an accelerated expansion phase is
$$\tau=\int_{\mathcal{H}_{B}}^{\mathcal{H}_{4}}\frac{d\mathcal{H}}{\dot{\mathcal{H}}_{+}}
+\int_{\mathcal{H}_{4}}^{\mathcal{H}_{5}}\frac{d\mathcal{H}}{\dot{\mathcal{H}}_{-}}=\frac{\sqrt{6}}{2\sqrt{\beta}}.$$
At that time, the universe ends its early accelerating expansion (inflation) and gracefully exits to a decelerated FRW expansion phase. By determining the value of $\beta$, the time interval of the accelerated expansion phase can be obtained. Since all problems of the standard cosmology can be solved in the bounce scenario, there is no need for a long interval inflation.
\subsubsection{Decelerated expansion (FRW)}
By crossing the transition point $\mathcal{H}_{5}$, the universe continues its expansion phase but decelerating. This is an important phase, since the horizon expands allowing quantum fluctuations to reenter the horizon at the superhorizon scale and transform to classical fluctuations propagating as particles to produce the universe which we see today. Therefore, this era is an essential phase for the large structures and galaxies formation. Assuming that the present Hubble value to be $\mathcal{H}_{0}$, the time required after transition to reach the present time can be given by
$$\tau=\int_{\mathcal{H}_{5}}^{\mathcal{H}_{0}}\frac{d\mathcal{H}}{\dot{\mathcal{H}}_{-}}=
\frac{3\sqrt{6}\mathcal{H}_{0}-\sqrt{\beta}-\sqrt{\beta-36\mathcal{H}_{0}^{2}}}
{6\sqrt{\beta}\mathcal{H}_{0}}.$$\\
\subsubsection{The Minkowskian fate}
Following the phase portrait in Fig. \ref{Fig:phase-portrait}, the universe evolves towards a Minkowskian space. Since the time required to reach that point is infinite, the universe has no future singularity. This can be show by
$$\tau=\int_{\mathcal{H}_0}^{0}\frac{d\mathcal{H}}{\dot{\mathcal{H}}_{-}}\to \infty.$$
In conclusion, the unimodular Lorenz gauge fixing provides a nonsingular bouncing cosmology, where crossing the phantom divide line is valid through de Sitter fixed points of Type IV singularity. In addition, the Minkowskian point represents the origin and the fate of the universe. Remarkably, the quantum fluctuation formulation in the Minkowskian space is more concrete than de Sitter. However, the model cannot predict the late accelerating expansion.
\subsection{Unimodular $f(T)$ Model}\label{Sec:5.4}
In order to fulfill the unimodular constraint, we employ the Lagrange multiplier method \cite{Nassur:2016yhc,Bamba:2016wjm}
\begin{equation}\label{Unimodular-action}
{\cal S}=\int d^{4}x~ \left[ |e| \left(\frac{1}{2 \kappa^2}f(\mathcal{T})-\lambda\right)+\lambda \right]+{\cal S}_{m},
\end{equation}
where $S_{m}$ is the action of the matter. The variation of the action (\ref{Unimodular-action}) with respect to the vierbein gives
\begin{equation}\label{uni-field_eqns}
\frac{1}{e} \partial_\mu \left( e S_a^{\verb| |\mu\nu} \right) f_\mathcal{T}-e_{a}{^\lambda}  \mathcal{T}^\rho_{\verb| |\mu \lambda} S_\rho^{\verb| |\nu\mu}f_\mathcal{T}+S_a^{\verb| |\mu\nu} \partial_\mu \mathcal{T} f_\mathcal{TT}
+\frac{1}{4} e_a{^\nu} (f-\lambda)=\frac{{\kappa}^2}{2} e_{a}{^\mu} \mathfrak{T}_\mu{^\nu},
\end{equation}
where $f_\mathcal{T}:=\frac{df}{d\mathcal{T}}$ and $f_\mathcal{TT}:=\frac{d^{2}f}{d\mathcal{T}^2}$, and ${\mathfrak{T}_{\mu}}^{\nu}$ is the usual energy-momentum tensor of matter fields. Assume that the material-energy tensor is taken for a perfect fluid as in Eq. (\ref{matter}), where $u^{\mu}=a(\tau)^{-3}\delta^{\mu}_{0}$ is the fluid 4-velocity in the unimodular universe (\ref{Unimodular-metric}). Then, the energy-momentum tensor reads $\mathfrak{T}_\mu{^\nu}=\textmd{diag}\left(\rho_{m},-p_{m},-p_{m},-p_{m}\right)$, where $\rho_{m}$ and $p_{m}$ are the energy density and pressure of the fluid in its rest frame.
\subsubsection{Reconstructing unimodular $f(T)$}
Using the field equations (\ref{uni-field_eqns}) for the matter (\ref{matter}), the modified Friedmann equations for the unimodular $f(T)$-gravity read
\begin{eqnarray}
  \rho_{m} &=& \frac{1}{2\kappa^{2}}\left[f(\mathcal{T})-\lambda+12a(\tau)^{6}\mathcal{H}^{2}f_{\mathcal{T}}\right], \label{fieldEqs1}\\
p_{m} &=& -\frac{1}{2\kappa^{2}}\left[f(\mathcal{T})-\lambda+4a(\tau)^{6}\left(6\mathcal{H}^{2}+\dot{\mathcal{H}}\right)f_{\mathcal{T}}
-48a(\tau)^{12}\left(\dot{\mathcal{H}}+3\mathcal{H}^{2}\right)\mathcal{H}^{2}f_{\mathcal{TT}}\right].\label{fieldEqs2}
\end{eqnarray}
It is clear that the unimodular gravity of the general relativity is recovered by setting $f(\tau)=\mathcal{T}(\tau)=-6a(\tau)^{6}\mathcal{H}^{2}$. Noting that the unimodular Lorenz gauge fixing forces the scale factor to produce a bouncing cosmology. Therefore, it convenient to rewrite $f(\mathcal{T})$ in terms of the unimodular time coordinates as $f(\tau)$, this is to avoid any possible discomfort due to the differences of the phase portrait before and after the bounce. Using Eq. (\ref{Torsionscalar}), one can obtain the first derivative
$$f_{\mathcal{T}}=\frac{\dot{f}}{\dot{\mathcal{T}}}=-\frac{1}{12}\frac{\dot{f}(\tau)}{a(\tau)^{6}\mathcal{H}^{2}
\left(\dot{\mathcal{H}}+3\mathcal{H}^{2}\right)},$$
and similarly for the second derivate $f_{\mathcal{TT}}$. Thus, the field equations (\ref{fieldEqs1}) and (\ref{fieldEqs2}) can be simplified to
\begin{eqnarray}
    \rho_{m}&=&~\frac{1}{2\kappa^2}\left[f(\tau)-\lambda-\frac{\mathcal{H} \dot{f}(\tau)}{\dot{\mathcal{H}}+3\mathcal{H}^{2}}\right],\label{uni-dens}\\
\nonumber p_{m}    &=&\frac{-1}{2\kappa^2}\left[f(\tau)-\lambda+\frac{\left(\ddot{\mathcal{H}}+6\mathcal{H}\dot{\mathcal{H}}
\right)\dot{f}(\tau)-\left(\dot{\mathcal{H}}+3\mathcal{H}^{2}\right)\ddot{f}(\tau)}
{3\left(\dot{\mathcal{H}}+3\mathcal{H}^{2}\right)^{2}}\right],\\
&~&\label{uni-press}
\end{eqnarray}
Recalling that the matter fluid is assumed to follow the linear equation of state, we use $\gamma=1+\omega_{m}$, the matter continuity equation reads
\begin{equation}\label{uni-continuity}
    \dot{\rho}_{m}(\tau)+3(1+w_{m})\mathcal{H}\rho_{m}(\tau)=0,
\end{equation}
which can be integrated to
\begin{equation}\label{uni-dens1}
    \rho_{m}(\tau)=\rho_{B,m}\exp{\left[-3(1+w_{m})\int \mathcal{H} d\tau\right]},
\end{equation}
\begin{figure}
\centering
\includegraphics[scale=.3]{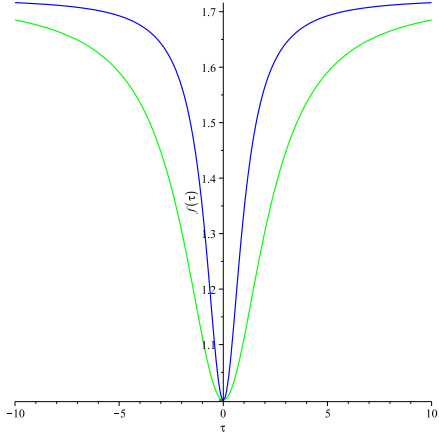}
\caption{The $f(\tau)$ function is symmetric about $\tau=0$, then Eq. (\ref{f(tau)}) produces the same $f(\mathcal{T})$ gravity before and after the bounce.}
\label{Fig:f(tau)}
\end{figure}
where the integration constant $\rho_{B,m}\equiv \rho(\tau_{B})$. We choose the bounce, $\mathcal{H}_{B}=0$, time to be at $\tau_{B}=0$, and therefore the maximum density at the bounce $\rho_{B,m}$ can be chosen to be less than Planck density in order to avoid the trans-Planckian problem. From Eqs. (\ref{uni-dens}) and (\ref{uni-dens1}), after some lengthy but direct manipulation, we evaluate a generic form of the unimodular $f(\tau)$ gravity as
\begin{equation}\label{generic.uni.f(tau)}
    f(\tau)=a(\tau)^3 \mathcal{H}\left[f_{0}-\int \frac{\lambda(\tau)+2\rho_{B,m} \kappa^2 a(\tau)^{-3(1+w)}}{a(\tau)^3 \mathcal{H}^2}\left(3{\mathcal{H}^2}+\dot{\mathcal{H}}\right) d\tau\right],
\end{equation}
where $f_{0}$ is a constant of integration. For the particular model at hand, we substitute (\ref{Uni-Hubble}) and (\ref{Uni-phase-portrait}) into (\ref{generic.uni.f(tau)}), then it reads
\begin{equation}\label{f(tau0)}
f(\tau)=\frac{\tau}{\sqrt{\beta \tau^{2}+1}}\left(f_0-\int \frac{\lambda(\tau)}{\tau^2 \sqrt{\beta \tau^{2}+1}} d\tau \right)
+\frac{2\rho_{B,m}\kappa^{2}}{\sqrt{\beta \tau^{2}+1}} \textmd{hypergeom}\left(\left[\frac{-1}{2},1+\frac{w_{m}}{2}\right],\left[\frac{1}{2}\right],-\beta \tau^{2}\right),
\end{equation}
Since the coefficient of $f_{0}\propto \sqrt{\mathcal{T}}$ and do not contribute in the field equations, we may omit the $f_{0}$ term without affecting the generality of the solution. For a special choice of the Lagrange multiplier $\lambda(\tau)=\Lambda$, Eq. (\ref{f(tau0)}) reduces to
\begin{equation}\label{f(tau)}
f(\tau)=\Lambda+\frac{2\rho_{B,m}\kappa^{2}}{\sqrt{\beta \tau^{2}+1}}\textmd{hypergeom}\left(\left[\frac{-1}{2},1+\frac{w_{m}}{2}\right],\left[\frac{1}{2}\right],-\beta \tau^{2}\right).
\end{equation}
The plot of Eq. (\ref{f(tau)}) is given in Fig. \ref{Fig:f(tau)}. This shows that the function $f(\tau)$ is symmetric about the bounce time $\tau_B$, and therefore it produces the same $f(\mathcal{T})$ gravity in both branches.
\subsubsection{Physics of torsion fluid}
In order to investigate the torsion role in the cosmological evolution, it is convenient to transform the field equations from the matter frame to the effective frame, which provides the gravitational sector as additional terms to Einstein's equations. So we write the modified Friedmann equations (\ref{fieldEqs1}) and (\ref{fieldEqs2}) in the case of unimodular $f(T)$ gravity as
\begin{eqnarray}
\mathcal{H}^2& =& \frac{\kappa^2}{3} a(\tau)^{-6} \left( \rho_m+  \rho_\mathcal{T} \right)\equiv \frac{\kappa^2}{3} a(\tau)^{-6}\rho_{\textmd{eff}}, \label{MFR1}\\
2 \dot{\mathcal{H}} + 9\mathcal{H}^{2}&=& - \kappa^2 a(\tau)^{-6} \left(p_m+p_\mathcal{T}\right)\equiv - \kappa^2 a(\tau)^{-6} p_{\textmd{eff}}.\label{MFR2}
\end{eqnarray}
We eliminate the scale factor contribution in the above by defining the effective equation of state parameter
\begin{equation}\label{effective_EoS}
    w_{\textmd{eff}}\equiv \frac{p_{\textmd{eff}}}{\rho_{\textmd{eff}}}=-3-\frac{2}{3}\frac{\dot{\mathcal{H}}}{\mathcal{H}^{2}}=-1-\frac{2}{\beta\tau^2}.
\end{equation}
Recalling Eq. (\ref{Uni-phase-portrait}), we find that the effective equation of state parameter is a function of $\mathcal{H}$ only. Otherwise, the field equations will not form an autonomous system anymore. Hence, we find that the unimodular $f(\mathcal{T})$ cosmology is in comfort with the one-dimensional phase portrait analysis.

By comparing Eqs. (\ref{uni-dens}), (\ref{uni-press}), (\ref{MFR1}) and (\ref{MFR2}), we write the torsion density and pressure $\rho_{\mathcal{T}}$ and $p_{\mathcal{T}}$, respectively, as
\begin{eqnarray}
    \rho_{\mathcal{T}}&=&\frac{1}{2\kappa^2}\left(6 a(\tau)^6 \mathcal{H}^{2}-f(\tau)+\lambda+\frac{\mathcal{H} \dot{f}}{\dot{\mathcal{H}}+3\mathcal{H}}\right),\label{rhoT}\\
p_{\mathcal{T}}&=&\frac{-1}{\kappa^2}\left[2a(\tau)^6 \left(\dot{\mathcal{H}}+3\mathcal{H}^2\right)-\frac{\ddot{\mathcal{H}}+9\mathcal{H}\left(\dot{\mathcal{H}}+\mathcal{H}^2\right)}
{6\left(\dot{\mathcal{H}}+3\mathcal{H}^2\right)^2}\dot{f}+\frac{\ddot{f}}{6\left(\dot{\mathcal{H}}+3\mathcal{H}^2\right)}\right]-\rho_{\mathcal{T}},\label{pT}
\end{eqnarray}
are the torsion contributions to the energy density and pressure, respectively. At the GR limit $f(\mathcal{T})=\mathcal{T}$, one can show that $\rho_{\mathcal{T}}=-p_{\mathcal{T}}=\frac{\lambda}{2\kappa^2}$, when the Lagrange multiplier acts a cosmological constant and the standard unimodular Friedmann equations are recovered. Acquiring the reasonable assumption that the effective fluid is conservative, we write the continuity equation of the torsion fluid
\begin{equation}\label{continuity}
    \dot{\rho}_{\mathcal{T}}+3\mathcal{H}(\rho_{\mathcal{T}}+p_{\mathcal{T}})\equiv 0.
\end{equation}
We next investigate the dynamical contribution of the torsion gravity in realizing a healthy bouncing scenario. Using Eqs. (\ref{rhoT}) and (\ref{pT}), we can define the torsion equation of state parameter as
\begin{equation}\label{Tor_EoS}
w_{\mathcal{T}}\equiv \frac{p_\mathcal{T}}{\rho_\mathcal{T}}=-1+\frac{3\rho_{B,m} \kappa^2 (1+w_{m})(1+\beta \tau^2)^{\tfrac{1-w_{m}}{2}}+2 \beta}{3\rho_{B,m} \kappa^2 (1+\beta \tau^2)^{\tfrac{1-w_{m}}{2}}-\beta^2 \tau^2}.
\end{equation}
\begin{figure}
\centering
\includegraphics[scale=.3]{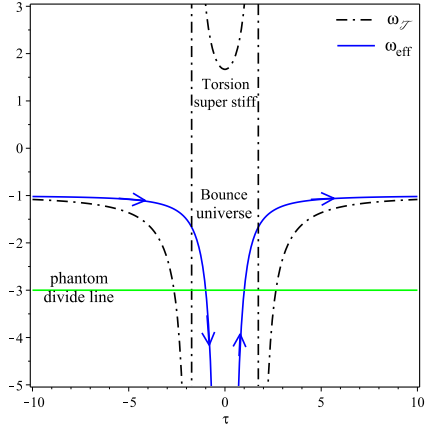}
\caption{The equation of state evolution of the effective and the torsion fluids.}
\label{Fig:effective-EoS}
\end{figure}
At bounce time $\tau_B$, the torsion fluid evolves as
\begin{equation}\label{bounce.limit}
    \lim_{\tau\to \tau_B} w_{\mathcal{T}}=w_{m}+\frac{2\beta}{3 \rho_{B,m} \kappa^2}> w_{m},
\end{equation}
while it is asymptotically evolves towards $w_{\mathcal{T}}\to -1$ as $\tau\to \pm \infty$. In order to be consistent with the stiff matter scale factor (\ref{uni.scale-factor1}), we take $w_{m}=1$. Then, Eq. (\ref{Tor_EoS}) reduces to
\begin{equation}\label{Tor_EoS_stiff}
    w_{\mathcal{T}}=-1+\frac{2(3\rho_{B,m} \kappa^2+\beta)}{3\rho_{B,m} \kappa^2 - \beta^2 \tau^2}.
\end{equation}
It is clear that the torsion equation of state is irregular at
$$\tau=\pm \frac{\kappa \sqrt{3\rho_{B,m}}}{\beta}.$$
The plot of Eq. (\ref{Tor_EoS_stiff}) in Fig. \ref{Fig:effective-EoS}, shows that the torsion fluid is asymptotically matches the effective equation of state, where both evolve towards $-1$. On the contrary, they have different behaviors about the bounce time, the universe effectively run deep into phantom regime as $w_\textmd{eff}\to -\infty$ as expected in bouncing scenarios. However, the torsion counterpart of $f(T)$ gravity has a large positive equation of state about the bounce time, i.e. $w_{{\mathcal{T}}} > 1$, this feature is important to solve the anisotropy growing problem during the contraction phase, which usually faces the bouncing cosmological models in the GR \cite{Kunze:1999xp,Xue:2011nw}. Also, the dynamical contribution of the torsion fluid allows to violate the null energy condition during the bounce only effectively, where the ordinary matter is kept safe, $0\geq w_m \geq 1$, as required by the stability and causality conditions. Additionally, the torsion fluid allow the model to avoid the ghost instability problem in the bouncing cosmology \cite{Novello:2008ra,Battefeld:2014uga}. Therefore, the model can perform a healthy bouncing scenario.
\subsubsection{Observational constraints}
In the above section, the qualitative analysis of the phase portrait has shown that the Lorentz gauge fixing of the unimodular teleparallel gravity enforces the FRW model to produce a nonsingular bounce scenario, where $-\infty < \tau < \infty$. In order to confront the model to some observational bounds, let us give some numerical estimates which help to obtain a reliable cosmology. We may choose the zero time $\tau=0$ at the bouncing point ($H=0$). From (\ref{Uni-Hubble}), we determine the constant $\tau_{B}=0$. For fixing $a_{B}$ and $\beta$, we take the conditions: (i) we normalize the scale factor (\ref{uni.scale-factor1}) to the present time $\tau=4.35\times 10^{17}$ s, (ii) we take the transition from acceleration to FRW deceleration at $t=10^{-32}$ s. These give $a_{B}=2.66\times 10^{-17}$ and $\beta=1.50\times 10^{64}$ s$^{-2}$. So it is not difficult to find that $|\mathcal{H}|=1.34\times 10^{7}$ GeV at de Sitter phases, which gives a good estimate of the energy scale to enter the following FRW era. In addition, the expansion of the phase portrait (\ref{Uni-phase-portrait}) about the Minkowski fixed point gives $\dot{\mathcal{H}}=-3\mathcal{H}^{2}$ which matches exactly the standard cosmology phase portrait during the FRW decelerated expansion phase. This guarantees a successful thermal history just as in the hot big bang scenario. Therefore, the model on its background level provides a good alternative to the standard cosmology (inflation + thermal history).

On the perturbation level, it has been shown that the bouncing cosmology can predict a scale invariant power spectrum of the scalar fluctuations as confirmed by observations. The scale invariant power spectrum using a single scalar with non-standard kinetic term has been investigated in Ref.~\cite{Cai:2012va}, also an extended study using two matter fields is given in Ref.~\cite{Cai:2013kja}. Possible mechanisms for generating a red tilt primordial curvature perturbations have been proposed in Ref.~\cite{Cai:2014bea}. Searching for clues to possible new physical parameters in the CMB observations at low redshifts has been investigated in Ref.~\cite{Cai:2015vzv}, the proposed investigation can be used to differential between different matter bouncing models. Also, small tensor-to-scalar ratio in bouncing cosmology can be achieved \cite{Cai:2016hea}. Another important parameter on the perturbation level of the model is the square of the sound speed $c_{s}^2$ of the scalar fluctuations. The valid range is $0\leq c_{s}^2 \leq 1$ to keep the stability and causality of the cosmic evolution. Since $c_{s}^2$ of the scalar fluctuations in $f(T)$ is dynamical \cite{Cai:2011tc}, it is essential to examine the validity of the model via the propagation of the sound speed of the scalar fluctuations. Otherwise, if $c_{s}^2$ becomes negative, the model will suffer from gradient instability. Also, if it exceeds the unity, the evolution will be acausal.

The above mentioned tests on the perturbative level of the theory need to be investigated in our model. However, we find that these are beyond the scope of this work. So we leave them for future work.
\section{Concluding remarks}\label{Sec:6}
In the teleparallel geometry, there are three fundamental diffeomorphism invariants. It has been shown that for a particular combination one can construct a teleparallel torsion invariant equivalent to the Ricci invariant up to a total derivative term. Since, the Ricci is also a Lorentz invariant but the total derivative is not, the torsion invariant is consequently not a Lorentz invariant. This serious issue forms a barrier to extend the teleparallel gravity reserving the Lorentz invariance. In this paper we investigate a simple idea to avoid this problem by fixing the total derivative term to a constant value. In this case we expect this term besides Ricci to be Lorentz invariants, thence we expect the same for the teleparallel torsion invariant.

We apply the procedure in the case of the FRW model, it results in a $\Lambda$CDM model. This leads to interpret the constant as the cosmological constant. The natural question then, why do Ricci and teleparallel torsion invariants almost compensate each other to keep the Lorentz symmetry? We study the stability of the obtained solution by analyzing its phase portrait. The analysis clarifies that the solutions give rise to big bang, bounce, turnaround or $\Lambda$CDM as separate cosmological models. We show that the $f(T)$ theories are in comfort with the phase portrait analysis, whereas its Friedmann system represents a one dimensional autonomous system. Then, we evaluate the $f(T)$ theory which generates these phase portraits.

In the above case, the obtained models are separated by fixed points so crossing from one to another is impossible. We find that the bounce and the standard cosmology can be unified in a single model, if we turn to the unimodular coordinates. We analyze the corresponding phase portrait. We show that the universe is nonsingular with an origin at Minkowski space, which has a concrete perturbation analysis. Also, we show that crossing between phantom and nonphantom regimes is valid through Type IV finite time singularities associated with the de Sitter phases. Moreover, the model matches the standard cosmology phase portrait after short inflation, so the hot big bang successes are recovered. The universe finally evolves towards a Minkowskian fate in an infinite time.

We employ the Lagrange multiplier method to formulate the unimodular $f(T)$ gravity which generates the corresponding phase portrait. Our investigation show that the torsion fluid has an essential role to avoid two main problems that usually face the bouncing models. The first is avoidance of forming ghost instabilities in the matter component by violating the null energy condition as in the general relativity or in the effective field theory. But, for the theory at hand, the violation is only in the torsion gravitational sector. The second is avoidance of forming anisotropic universe in the contraction period before bouncing up to expansion. This can be avoided if the matter component becomes super stiff $w_{m}\gg 1$ with a minimal value greater then the matter component, in order to have a growth rate much greater than the anisotropy. But the super stiff ad hoc condition violates the causality. In this theory, we find that the torsion fluid runs with the the effective fluid in general, except about the bounce point, the effective fluid runs deep into phantom regime as expected, while the torsion fluid becomes super stiff about the bounce point, particularly during $|\tau| < \frac{\kappa \sqrt{3\rho_{B,m}}}{\beta}$, which is in agreement with $\beta$ being that an indicator of how fast bounce occurs such that the bounce is fast as large as $\beta$. In conclusion, the higher order torsion gravity plays a crucial role to perform a healthy bouncing cosmology.
\section*{Acknowledgments}
This work is partially supported by the Egyptian Ministry of Scientific Research under project No. 24-2-12.
%
\end{document}